\documentclass[twocolumn,10pt,conference]{IEEEtran}
\normalsize

\ifCLASSINFOpdf
\else
\fi

\usepackage{longtable}
\usepackage{amsmath}
\usepackage{cancel}
\usepackage{amssymb}
\usepackage{graphicx}
\usepackage{longtable}
\usepackage{multirow}
\usepackage{array}
\usepackage[labelsep=period]{caption}
\usepackage{setspace}
\usepackage{float}
\usepackage{subfig}
\usepackage{cite}
\usepackage{flushend}
\usepackage{cleveref}
\newenvironment{definition}[1][Definition]{\begin{trivlist}
		\item[\hskip \labelsep {\bfseries #1}]}{\end{trivlist}}

\newcommand{\qed}{\nobreak \ifvmode \relax \else
	\ifdim\lastskip<1.5em \hskip-\lastskip
	\hskip1.5em plus0em minus0.5em \fi \nobreak
	\vrule height0.75em width0.5em depth0.25em\fi}

\begin{document}
%
\title{Outage Probability of the EH-based Full-Duplex AF and DF Relaying Systems in $\alpha$-$\mu$ Environment }

\author{Galymzhan Nauryzbayev, Mohamed Abdallah$^{\ddagger}$, and Khaled M. Rabie$^*$,\\
	\IEEEauthorblockA{Department of Radio Engineering, Electronics and Telecommunications, \\
	LN Gumilyov Eurasian National University, Astana, 010005, Kazakhstan\\
		$^{\ddagger}$Division of Information and Computing Technology, College of Science and Engineering, \\
		Hamad Bin Khalifa University, Qatar Foundation, Doha, Qatar \\
		$^*$School of Engineering, Manchester Metropolitan University, Manchester, UK, M15 6BH,\\
		Email: nauryzbayevg@gmail.com; $^{\ddagger}$moabdallah@hbku.edu.qa; $^*$k.rabie@mmu.ac.uk}}

\maketitle

\thispagestyle{empty}

\begin{abstract}
Wireless power transfer and energy harvesting have attracted a significant research attention in terms of their application in cooperative relaying systems. Most of existing works in this field focus on the half-duplex (HD) relaying mechanism over certain fading channels, however, in contrast, this paper considers a dual-hop full-duplex (FD) relaying system over a generalized independent but not identically distributed $\alpha$-$\mu$ fading channel, where the relay node is energy-constrained and entirely depends on the energy signal from the source node. Three special cases of the $\alpha$-$\mu$ model are investigated, namely, Rayleigh, Nakagami-$m$ and Weibull fading. As the system performance, we investigate the outage probability (OP) for which we derive exact unified closed-form expressions. The provided Monte Carlo simulations validate the accuracy of our analysis. Moreover, the results obtained for the FD scenario are compared to the ones related to the HD. The results demonstrate that the decode-and-forward relaying outperforms the amplify-and-forward relaying for both HD and FD scenarios. It is also shown that the FD scenario performs better than the HD relaying systems. Finally, we analyzed the impact of the fading parameters $\alpha$ and $\mu$ on the achievable OP. 
\end{abstract}


\IEEEpeerreviewmaketitle

\section{Introduction}
\IEEEPARstart{W}{ireless} power transfer has attracted considerable research interest as a potential candidate able to extend the lifespan of wireless battery-powered devices \cite{L2,K1,GS2}. Exploiting radio-frequency (RF) signals for simultaneous energy and information delivery, best known as simultaneous wireless information and power transfer (SWIPT), is believed to be one of the most efficient techniques for wireless energy-harvesting (EH). There are three prominent architectures for SWIPT systems in the literature, namely, ideal relaying, power-splitting (PS) and time-switching (TS) protocols \cite{L4,K2,L1,GS3,L6,khaled1}.

In SWIPT systems, the relaying node uses the received RF signal as a source of energy and then utilizes it to forward the intended data to a corresponding destination. The performance of two-hop amplify-and-forward (AF) relaying systems over Rayleigh fading authors was studied in \cite{L4}, where the authors considered three relaying protocols: time-switching relaying (TSR), ideal relaying receiver (IRR) and power-splitting relaying (PSR). In \cite{L5}, the authors estimated the outage probability (OP) for a two-hop decode-and-forward (DF) underlay cooperative cognitive network deploying the TSR and PSR protocols, again, in Rayleigh fading channels. Furthermore, accurate numerical expressions of the achievable throughput and ergodic capacity of a DF relaying system deploying the TSR and PSR protocols over Rayleigh channels were derived in \cite{L6}. Moreover, the OP in dual-hop AF and DF relaying networks was studied in \cite{khaled1,khaled3,L8} considering both full-duplex (FD) and half-duplex (HD) with several EH protocols. In addition, the authors in \cite{L4,GS3} and \cite{khaled2} focused on AF relaying systems with EH constraints for an IRR protocol. The transmission rate and OP for FD-DF relaying networks were investigated in \cite{EH_FD3} and \cite{EH_FD4}, respectively. 

Another aspects such as energy efficiency and security issues in a wireless power transfer (WPT) enabled FD-DF relaying network were studied in \cite{EH_FD1,icc_galym}. The authors in \cite{EH_MIMO3} and \cite{EH_MIMO1} investigated the secrecy rate and energy efficiency in wireless powered massive multiple-input multiple-output (MIMO) networks, respectively. In addition, the authors in \cite{EH_MIMO2} analyzed the degrading effect such as inter-relay interference in the WPT-enabled MIMO virtual FD relaying scheme. Recently, the authors in \cite{EH_NOMA1,EH_NOMA2,EH_NOMA3} considered a non-orthogonal multiple access (NOMA) approach in wireless powered relaying systems. For instance, the work in \cite{EH_NOMA1} and \cite{EH_NOMA2} investigated the outage and data rate performance of PS-based downlink cooperative SWIPT NOMA systems. Furthermore, the OP and energy efficiency of WPT-based NOMA-inspired AF systems over Nakagami-$m$ fading were studied in \cite{EH_NOMA3}.

Very recently, the author in \cite{badarneh} analyzed the OP in wireless powered HD-DF relaying systems over $\alpha$-$\mu$ fading channels. However, to the best of our knowledge, the performance analysis of wireless powered FD-AF relaying networks over $\alpha$-$\mu$ fading have not appeared in the literature. Therefore, we dedicate the paper to analyzing the OP over independent and not necessarily identically distributed (i.n.i.d.) $\alpha$-$\mu$ fading channels in a dual-hop TSR-based FD-AF/DF relaying network. It is also worthwhile noting that small-scale fading channels, such as Nakagami-$m$, Rayleigh, etc., can be described by a generalized $\alpha$-$\mu$ distribution \cite{magableh}. The derived expressions are unified in the sense that they represent various fading channels which are obtainable from the $\alpha$-$\mu$ statistical model. Moreover, the results are compared with those of the DF/AF-HD relaying system. Results reveal that the performance of DF relaying systems is superior to that of the AF ones while the FD networks achieve better performance than the HD ones. Throughout this work, our theoretical analysis is validated by Monte Carlo simulations. 

The remainder of this paper is structured as follows. Section II describes the system model and defines the performance metric adopted in this paper. Sections III derives new analytical expressions for the OP over i.n.i.d. $\alpha$-$\mu$ fading channels for the TSR protocols. Analytical and simulated results are provided and discussed in Section IV. Finally, Section V concludes the paper and outlines the main findings of this work.

\section{System and Channel Model}
\begin{figure}[!t]
	\centering
	\includegraphics[width=0.7\columnwidth]{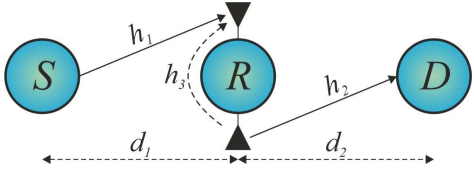}
	\caption{Diagram of the considered two-hop FD relaying system.}
	\label{system}
\end{figure}

The system model considered here consists of three nodes: a source ($S$), a relay ($R$) and a destination ($D$) as illustrated in Fig. \ref{system}. The overall $S$-$to$-$D$ communication is realized over two time periods. The first phase is dedicated for the EH and $S$-$to$-$R$ transmission while the second phase is used for the $R$-$to$-$D$ transmission when $R$ can operate either in the AF or DF mode. We also assume that $R$ is not equipped with external power supply, but is deployed with two antennas that allows it to simultaneously receive and transmit information. We assume that the amount of power used by $R$ during information processing is negligible. During the first phase, $R$ scavenges the energy from the signal transmitted by $S$ with power $P_S$. Additionally, no direct link is assumed between $S$ and $D$ and that all nodes, except $R$, are equipped with a single-antenna operating in the FD mode. The $S$-$to$-$R$ and $R$-$to$-$D$ links, represented by $h_1$ and $h_2$, follow quasi-static i.n.i.d. $\alpha$-$\mu$ fading. $d_1$ and $d_2$ represent the $S$-$to$-$R$ and $R$-$to$-$D$ distances, respectively; the corresponding path-loss exponents are given by $m_1$ and $m_2$. Note that the fading coefficients vary independently from one transmission block time $T$ to another.

Since we assumed that the channel envelope $(r)$ follows the $\alpha$-$\mu$ distribution, the probability density function (PDF) of the $i-$th hop is given by \cite{galym}
\begin{equation}
\label{pdf_h_i}
f_{h_i}(r) = \frac{\alpha_i \mu_i^{\mu_i} r^{\alpha_i \mu_i - 1}}{\hat{r}^{\alpha_i\mu_i} \Gamma(\mu_i)} \exp\left( -\frac{\mu_i}{\hat{r}^{\alpha_i}} r^{\alpha_i} \right),
\end{equation}
where $\hat{r}$ indicates the $\alpha_i-$root mean value given by $\hat{r} = \sqrt[\alpha_{i}]{\mathbb{E}\left[r^{\alpha_i}\right]}$, $\alpha_i > 0$ is an arbitrary parameter,  $\mathbb{E}\left[\cdot\right]$ is the expectation operator and $\Gamma\left(\cdot\right)$ denotes the Gamma function defined as $\Gamma(s) = \int_{0}^{\infty} t^{s-1} e^{-t} \textrm{d}t $ \cite[Eqs. (8.310) and (8.350.3)]{gradstein}. Also, $\mu_i\ge \frac{1}{2}$ is the inverse of normalized variance of $r^{\alpha_i}$ given by
\begin{equation}
\mu_i = \frac{\mathbb{E}\left[r^{\alpha_i}\right]}{\left( \mathbb{E}\left[ r^{2\alpha_i} \right] - \mathbb{E}^2\left[r^{\alpha_i}\right] \right)}.
\end{equation}

Note that the $\alpha$-$\mu$ distribution is the most appropriate fading statistical model that describes small-scale fading channels such as Rayleigh ($\alpha=2$, $\mu=1$), Weibull ($\alpha$ is the fading parameter with $\mu=1$), Nakagami$-m$ ($\mu$ is the fading parameter with $\alpha=2$), etc. \cite{magableh}.

The OP can be expressed as
\begin{align}
\label{Pout_gen}
P_{\textrm{out}} = \textrm{Pr}\left(C^{\textrm{AF/DF}} < \mathcal{R}\right) = \textrm{Pr}\left(\gamma_{\textrm{eff}}^{\textrm{AF/DF}} < \nu \right),
\end{align} 
where $\mathcal{R}$ and $C$ indicate the minimum required rate for the TSR protocol and the instantaneous capacity of the effective end-to-end signal-to-noise ratio (SNR) $\gamma_{\textrm{eff}}$, respectively. $\nu = 2^{\frac{\mathcal{R}}{1-\eta}} - 1$ denotes the threshold SNR supporting $\mathcal{R}$ for the FD relaying systems.
\begin{figure}[!t]
	\centering
	\includegraphics[width=0.9\columnwidth]{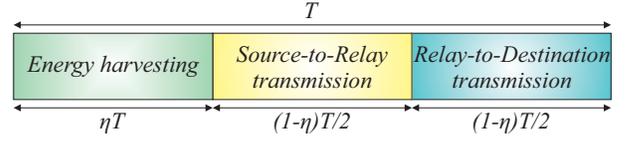}
	\caption{Time frame structure for the TSR relaying system.}
	\label{Protocols}
\end{figure}

\section{Outage Probability Analysis}
In this section, we analyze the OP of the two-hop time-switching EH-based FD system with DF and AF relaying. The given time $T$ required for $S$-$to$-$D$ information transmission is formed by three consecutive time slots (TSs). The first TS ($\eta T$) is dedicated for EH while the remaining two TSs (each equal to $(1-\eta)T/2$) are designated to maintain the $S$-$to$-$R$ and $R$-$to$-$D$ data transmissions, where the EH time factor is defined by $0 \le \eta \le 1$ as shown in Fig. \ref{Protocols}.

The received signal at $R$ can be written as \cite{khaled1}
\begin{equation}
\label{relay_signal}
y_R(t) = \sqrt{\frac{P_S}{d_1^{m_1}}} h_1 s(t) + n_a(t),
\end{equation}
where $P_S$ stands for the source transmit power. The information signal and the noise at $R$ are denoted as $s(t)$, satisfying $\mathbb{E}\left[ |s(t)|^2 \right] = 1$, and $n_a(t)$, with variance $\sigma_a^2$, respectively. Accordingly, the energy harvested at $R$ can be presented as  
\begin{equation}
\label{harv_energy}
E_H = \theta \eta T \left( \frac{P_S}{d_1^{m_1}} h_1^2 + \sigma_a^2 \right), 
\end{equation}
where the EH conversion efficiency $\theta$ $(0<\theta\le1)$ is mainly affected by the circuitry. The energy harvested from the noise term can be neglected due to its insignificance. Therefore, we can express the relay transmit power as
\begin{equation}
\label{relay_power}
P_R = \frac{E_H}{(1-\eta)T} = \frac{\theta \eta P_S h_1^2}{(1-\eta) d_1^m} = \frac{\kappa P_S h_1^2}{d_1^{m_1}},
\end{equation}
where $\kappa = \frac{\theta \eta}{1-\eta}$.

The received signal within the information transmission mode can be given by
\begin{equation}
\label{relay_IT}
y_R(t) = \sqrt{\frac{P_S}{d_1^{m_1}}} h_1 s(t) + h_3 r(t) + n_a(t),
\end{equation}
where $r(t)$ and $h_3$ denote the loop-back interference (LBI) signal (due to the FD mode, with $\mathbb{E}\left[|r(t)|^2\right] = P_R$) and LBI channel, respectively. 

Note that in the FD mode, $R$ applies interference cancellation to reduce LBI since it knows its own signal. Thus, the post-canceled received signal at $R$ can be presented as 
\begin{equation}
\label{relay_IT1}
y_R(t) = \sqrt{\frac{P_S}{d_1^{m_1}}} h_1 s(t) + \hat{h_3}\hat{r}(t) + n_a(t),
\end{equation}
where $\hat{h_3}$ indicates the residual LBI channel given by the interference cancellation impairments with $\mathbb{E}\left[|\hat{r}(t)|^2\right] = P_R$.

\subsection{FD with DF Relaying}
For the DF relaying, the decoded source signal $\bar{s}(t)$ is forwarded by $R$ to $D$; hence, the received signal at $D$ can be expressed as 
\begin{equation}
\label{yd_df}
y_D(t) = \sqrt{\frac{P_R}{d_2^{m_2}}} h_2 \bar{s}(t) + n_D(t).
\end{equation}

Using \eqref{relay_power}, \eqref{relay_IT1} and \eqref{yd_df}, the SNR values at $R$ and $D$ can be accordingly written as
\begin{align}
\label{gammaR_df}
\gamma_R &= \frac{P_S h_1^2}{P_R d_1^{m_1} h_3^2} = \frac{1}{\kappa h_3^2},\\
\label{gammaD_df}
\gamma_D &= \frac{P_R h_2^2}{d_2^{m_2} \sigma_D^2} = \frac{\kappa P_S h_1^2 h_2^2}{d_1^{m_1} d_2^{m_2} \sigma_D^2}.
\end{align}

The instantaneous capacity of the $S$-$to$-$D$ link can be written as
\begin{align}
\label{C_DF}
C^{\textrm{DF}} &= (1 - \tau) \log_2 (1 + \gamma_{\textrm{eff}}^{\textrm{DF}}) \nonumber\\
&= (1 - \tau) \log_2 (1+\min(\gamma_R,\gamma_D)).
\end{align}

Now, substituting \eqref{C_DF} into \eqref{Pout_gen} and assuming $Z = X Y$ $(X = h_1^2, Y = h_2^2)$ and $V = h_3^2$, the OP can be determined as 
\begin{equation}
P_{\textrm{out}}^{\textrm{DF}} = \textrm{Pr}\left(\min\left( \frac{1}{\kappa V}, \frac{\kappa P_S Z}{d_1^{m_1} d_2^{m_2} \sigma_D^2} \right)< \nu \right).
\end{equation}

Since the random variables (RVs) $V$ and $Z$ are independent, the corresponding OP can be expressed as
\begin{equation}
\label{Pout_DF}
P_{\textrm{out}}^{\textrm{DF}} = 1 - F_{V}\left(\frac{1}{\kappa \nu}\right) \left(1 - F_{Z}\left(\nu \frac{d_1^{m_1} d_2^{m_2} \sigma_D^2}{\kappa P_S}\right) \right),
\end{equation}
where $F_V(r) = \int_{0}^{r} f_V(z) \textrm{d}z$ and $F_{Z}(r) = \int_{0}^{r} f_Z(z) \textrm{d}z$ are the cumulative distribution functions (CDFs) of RVs $V$ and $Z$, respectively. 

\begin{definition}
	Let $R$ be a continuous RV with generic PDF $f(r)$ defined over the support $t_1 < r < t_2$ and $Q = g(R)$ be an invertible function of $R$ with inverse function $R = \nu\left( Q \right)$. Then, using the \textquotedblleft change of variable\textquotedblright~method, the PDF of $Q$ can be expressed as 
	\begin{equation}
	\label{change}
	f_{Q}(q) = f_{R}\left( \nu(q) \right)|\nu'\left(q\right)|
	\end{equation}
	defined over the support $g(t_1) < q < g(t_2)$.
\end{definition}

The PDFs of $X$, $Y$ and $V$ can be rewritten using the definition given by \eqref{change} as
\begin{align}
\label{pdf_h_i_2}
f_{h_i^2}(r) &= \frac{\alpha_i \mu_i^{\mu_i} r^{\frac{\alpha_i \mu_i}{2} - 1}}{2\hat{r}^{\alpha_i\mu_i/2} \Gamma(\mu_i)} \exp\left( -\frac{\mu_i}{\hat{r}^{\alpha_i}} r^{\frac{\alpha_i}{2}} \right).
\end{align}

Using \eqref{pdf_h_i_2}, the CDF of $V$ can be easily obtained as
\begin{equation}
\label{cdf_V}
F_V = \frac{\gamma_{inc}\left(\mu_3, \frac{\lambda_3}{(\kappa \nu)^{\alpha_3/2}} \right)}{\Gamma(\mu_3)},
\end{equation}
where $\gamma_{inc}(s,x) = \int_{0}^{x} t^{s-1}\exp(-t) \textrm{d}t$ indicates the lower Gamma function \cite[Eq. (8.350.1)]{gradstein}.

Next, we define the PDF of $Z=XY$ as follows
{\allowdisplaybreaks
\begin{align}
f_{Z} (r) &= \int_{0}^{\infty} \frac{f_{X}(u) f_{Y}\left(\frac{r}{u}\right)}{|u|} \textrm{d}u\nonumber\\
& = \frac{\alpha_2 \lambda_1^{\frac{3\mu_1 - \mu_2}{2}} \lambda_2^{\frac{3\mu_2 - \mu_1}{2}} r^{\frac{\alpha_2 (\mu_1 + \mu_2)}{4} - 1} }{\Gamma(\mu_1)\Gamma(\mu_2)} \nonumber\\
&~~~\times K_{\mu_1-\mu_2}\left( 2\sqrt{\lambda_1 \lambda_2 r^{\alpha_2}} \right),
\end{align}}when $\alpha_1=\alpha_2$ and $\lambda_i = \frac{\mu_i}{\hat{r}^{\frac{\alpha_i}{2}}}$. $K_{n}(\cdot)$ is the $n$-th order modified Bessel of the second kind. Therefore, the CDF of $F_{Z}(r)$ can be determined using \cite[Eq. (14) and (26)]{adamchik} as in \eqref{cdf_Z}, as shown at the top of the next page. 

After substituting \eqref{cdf_V} and \eqref{cdf_Z} into \eqref{Pout_DF}, we can finally provide a closed-form expression of the FD-DF system based OP as in \eqref{Pout_DF1}, as shown at the top of the next page.

\begin{figure*}[!t]
	\normalsize
	\begin{align}
	\label{cdf_Z}
	F_{Z}(r) = \frac{\lambda_1^{\frac{3\mu_1 - \mu_2}{2}} \lambda_2^{\frac{3\mu_2 - \mu_1}{2}} }{\Gamma(\mu_1) \Gamma(\mu_2)} \left( \frac{\nu \left(\beta_3 r + \beta_4\right)}{\beta_1 - \beta_2 \nu r} \right)^{\frac{\alpha_2(\mu_1 + \mu_2)}{4}}  G_{1,3}^{2,1}\left( \left. \lambda_1 \lambda_2 \left( \frac{\nu \left(\beta_3 r + \beta_4\right)}{\beta_1 - \beta_2 \nu r} \right)^{\frac{\alpha_2}{2}} \right\vert \begin{array}{c}
	1-\frac{\mu_1 + \mu_2}{2}\\
	\frac{\mu_1 - \mu_2}{2}, -\frac{\mu_1 + \mu_2}{2}, \frac{\mu_2 - \mu_1}{2}
	\end{array}\right)
	\end{align}
	\hrulefill
	\begin{multline}
	\label{Pout_DF1}
	P_{\textrm{out}}^{\textrm{DF}} = 1- \frac{\gamma_{inc}\left(\mu_3, \lambda_3 \frac{1}{(\kappa \nu)^{\alpha_3/2}} \right)}{\Gamma(\mu_3)} \left(1 - \frac{\lambda_1^{\frac{3\mu_1 - \mu_2}{2}} \lambda_2^{\frac{3\mu_2 - \mu_1}{2}} }{\Gamma(\mu_1) \Gamma(\mu_2)} \left( \frac{\nu d_1^{m_1} d_2^{m_2} \sigma_R^2 \left( \frac{\nu d_1^{m_1} d_2^{m_2} \sigma_D^2}{P_S} + 1 \right)}{\kappa P_S - \kappa\nu^2 d_1^{m_1} d_2^{m_2} \sigma_D^2 } \right)^{\frac{\alpha_2(\mu_1 + \mu_2)}{4}} \right. \\ 
	\left.\times G_{1,3}^{2,1}\left( \left. \lambda_1 \lambda_2 \left(\frac{\nu d_1^{m_1} d_2^{m_2} \sigma_R^2 \left( \frac{\nu d_1^{m_1} d_2^{m_2} \sigma_D^2}{P_S} + 1 \right)}{\kappa P_S - \kappa\nu^2 d_1^{m_1} d_2^{m_2} \sigma_D^2 } \right)^{\frac{\alpha_2}{2}} \right\vert \begin{array}{c}
	1-\frac{\mu_1 + \mu_2}{2}\\
	\frac{\mu_1 - \mu_2}{2}, -\frac{\mu_1 + \mu_2}{2}, \frac{\mu_2 - \mu_1}{2}
	\end{array}\right) \right) 
	\end{multline}
	\hrulefill
\end{figure*}

\begin{figure*}[!t]
	\centering
	\subfloat[Rayleigh ($\alpha=2$ and $\mu=1$).]{
		\label{subfig:Rayleigh}
		\includegraphics[width=0.3\textwidth]{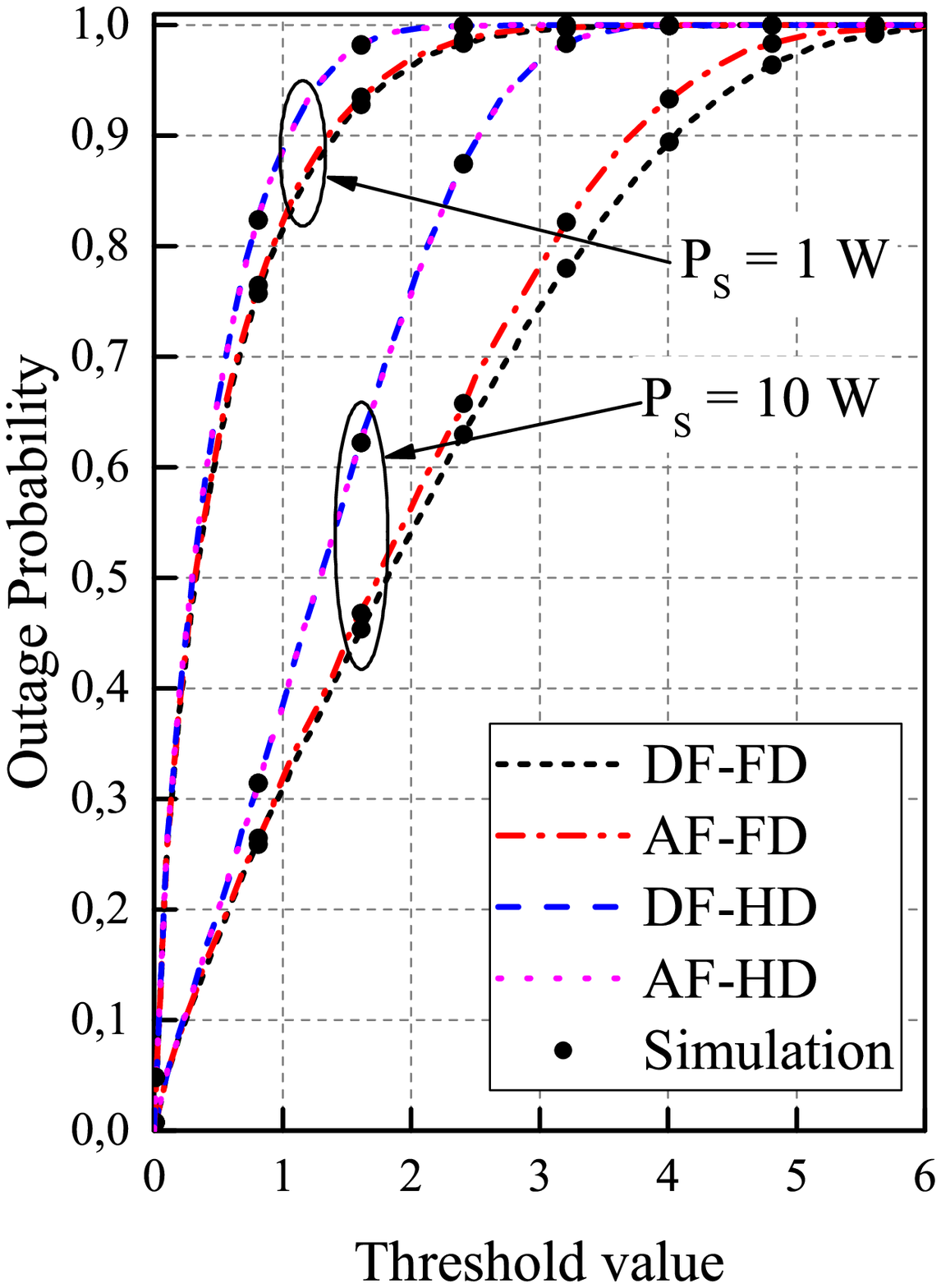}}
	\subfloat[Weibull ($\alpha=3$ and $\mu=1$).]{
		\label{subfig:Weibull}
		\includegraphics[width=0.3\textwidth]{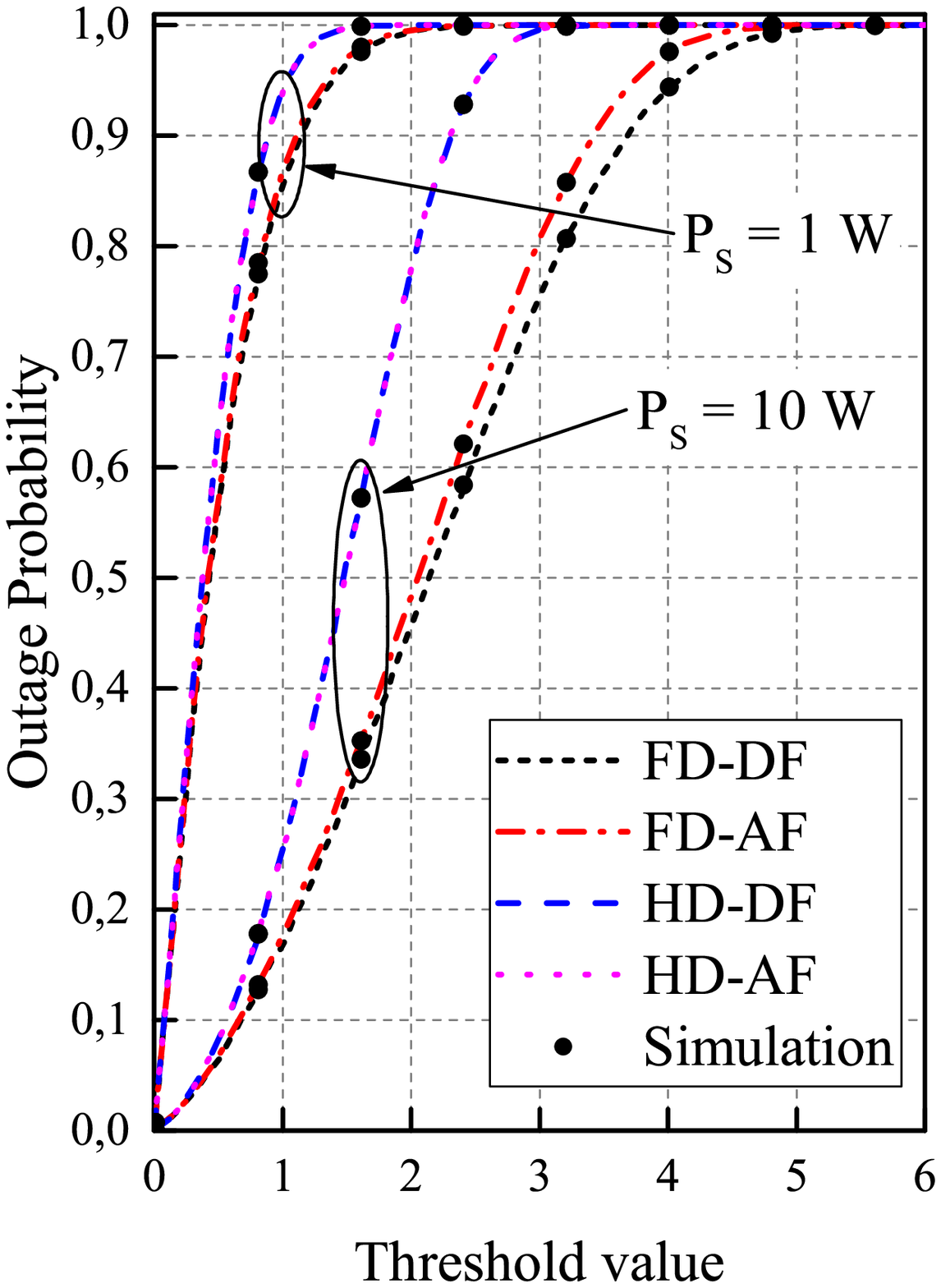}}
	\subfloat[Nakagami-$m$ ($\alpha = 2$ and $\mu=2$).]{
		\label{subfig:Nakagami}
		\includegraphics[width=0.3\textwidth]{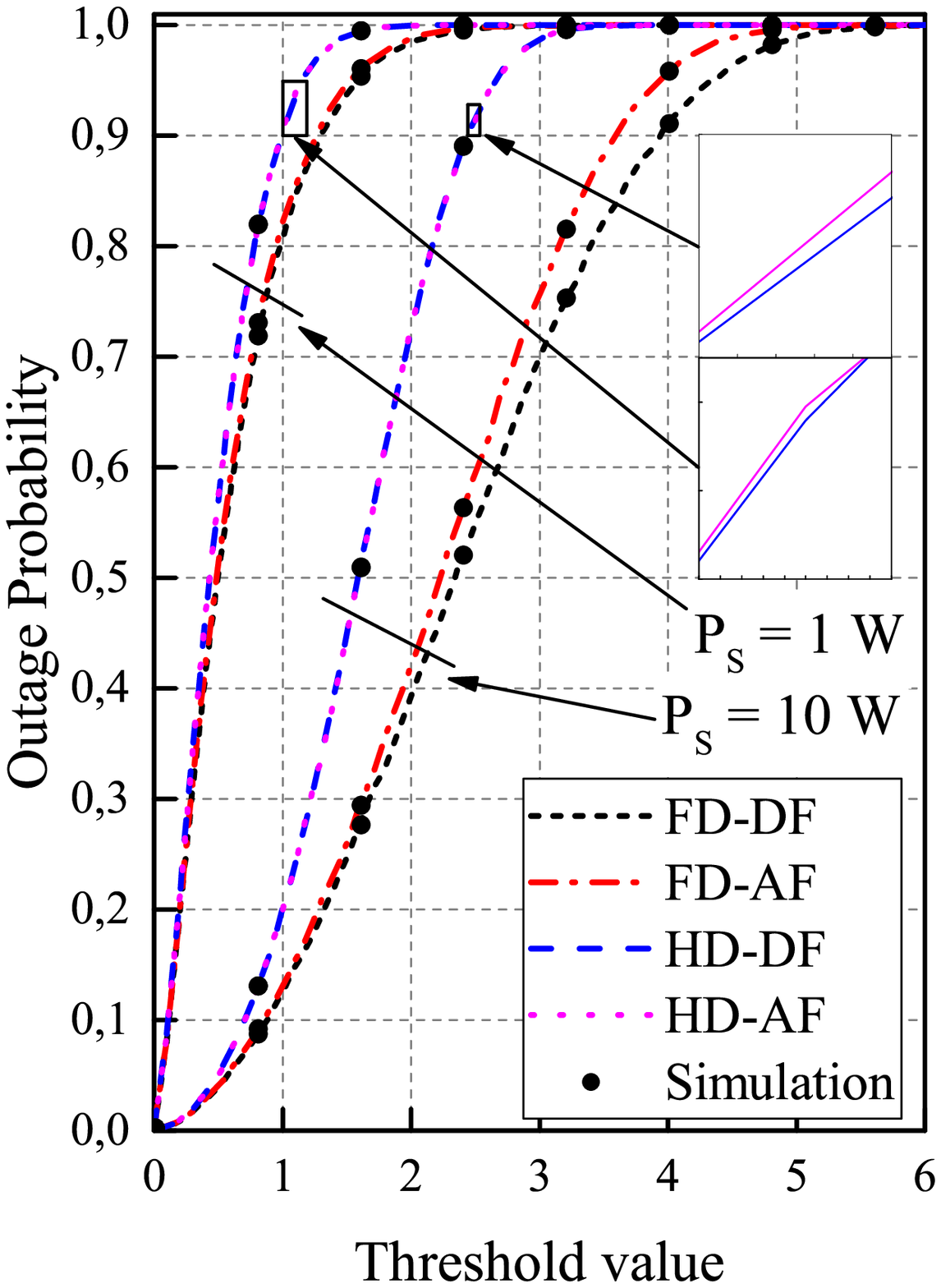}}
	\caption{OP versus $\mathcal{R}$ for both HD and FD relaying systems with various parameters of the $\alpha$-$\mu$ fading distribution.}
	\label{results1}
\end{figure*}

\subsection{FD with AF Relaying}
During the $R$-$to$-$D$ transmission, the relay node amplifies the received signal and then forwards it to the destination node. Therefore, the received signal at $D$ can be expressed as
\begin{align}
\label{rx_dest_signal}
y_D(t) &= \sqrt{\frac{P_S P_R}{d_1^{m_1} d_2^{m_2}}}G h_1 h_2 s(t) + \sqrt{\frac{P_R}{d_2^{m_2} }}G h_2 h_3 r(t)  \nonumber\\
&~~~+ \underset{\textrm{noise term}}{\underbrace{\sqrt{\frac{P_R}{d_2^{m_2}}}G h_2 n_a(t)+ n_D(t)}},
\end{align}
where $G = 1/\sqrt{\frac{P_S}{d_1^{m_1}} h_1^2 + P_R h_3^2 + \sigma_R^2}$ denotes the relay gain. $n_R(t) = n_a(t) + n_c(t)$ denotes the overall noise at $R$ with variance $\sigma_R^2 = \sigma_a^2 + \sigma_c^2$, where $n_c(t)$ indicates the noise caused by the information receiver while $n_D(t)$ indicates the noise term at $D$ with variance $\sigma_D^2$. 

With this in mind, after some algebraic manipulations, we can write the SNR at $D$ as 
\begin{equation}
\label{dest_snr_af}
\gamma_D = \frac{P_S h_1^2 h_2^2}{P_R d_1^{m_1} h_2^2 h_3^2 + \frac{P_S d_2^{m_2} h_1^2\sigma_R^2}{P_R} + h_3^2 d_1^{m_1} d_2^{m_2} \sigma_R^2}.
\end{equation}

After substituting \eqref{relay_power} into \eqref{dest_snr_af}, the SNR $\gamma_D$ can be re-expressed as follows
\begin{equation}
\label{snr_af}
\gamma_D = \frac{\beta_1 Z}{\beta_2 V Z + \beta_3 V  + \beta_4},
\end{equation}
where $\beta_1 = P_S$, $\beta_2 = \kappa P_S $, $\beta_3 = d_1^{m_1} d_2^{m_2} \sigma_R^2$ and $\beta_4 = \frac{d_1^{m_1} d_2^{m_2} \sigma_R^2}{\kappa}$.

The instantaneous end-to-end capacity can be expressed as
\begin{align}
C^{\textrm{AF}} &= (1 - \tau) \log_2 (1 + \gamma_{\textrm{eff}}^{\textrm{AF}}) \nonumber\\
&= (1 - \tau) \log_2 (1+\gamma_D).
\end{align}

Using \eqref{snr_af}, the OP given by \eqref{Pout_gen} can be expressed as
\begin{align}
P_{\textrm{out}}^{\textrm{AF}} &= \textrm{Pr}\left( \frac{\beta_1 Z}{\beta_2 V Z + \beta_3 V  + \beta_4} < \nu \right) \nonumber\\
&= \textrm{Pr}\left( Z < \frac{\nu\left( \beta_3 V + \beta_4 \right)}{\beta_1 - \beta_2 \nu V}\right). 
\end{align} 

The fact that $Y$ is a positive value means 
\begin{equation}
\label{Pout}
P_{\textrm{out}}^{\textrm{AF}} = \begin{cases}
\textrm{Pr}\left( Z < \frac{\nu\left( \beta_3 V + \beta_4 \right)}{\beta_1 - \beta_2 \nu V} \right), \quad \quad ~~ V<\frac{\beta_1 }{\nu \beta_2}; \\
\textrm{Pr}\left( Z > \frac{\nu\left( \beta_3 V + \beta_4 \right)}{\beta_1 - \beta_2 \nu V} \right) = 1,\quad V>\frac{\beta_1 }{\nu \beta_2}.
\end{cases}
\end{equation}
Therefore, the OP can be calculated as
\begin{align}
\label{P_out_AF}
P_{\textrm{out}}^{\textrm{AF}} = \int_{\frac{\beta_1 }{\nu \beta_2}}^{\infty} f_{V}(r) \textrm{d}r + \int_{0}^{\frac{\beta_1 }{\nu \beta_2}}  F_Z(r) f_{V}(r) \textrm{d}r.
\end{align}

Finally, after substituting \eqref{cdf_Z} into \eqref{P_out_AF} and some algebraic manipulation, the OP can be expressed as in \eqref{Pout_AF}, shown at the top of the next page. 

\begin{figure*}[!t]
	\normalsize
	\begin{align}
	\label{Pout_AF}
	P_{\textrm{out}}^{\textrm{AF}} &= 1- \frac{\gamma_{inc}\left(\mu_3, \lambda_3 \frac{1}{(\kappa \nu)^{\alpha_3/2}} \right)}{\Gamma(\mu_3)} + \frac{ \alpha_3 \lambda_1^{\frac{3\mu_1 - \mu_2}{2}} \lambda_2^{\frac{3\mu_2 - \mu_1}{2}} \lambda_3^{\mu_3}}{2 \Gamma(\mu_1) \Gamma(\mu_2) \Gamma(\mu_3)}	\int_{0}^{\frac{1}{\kappa \nu}} r^{\frac{\alpha_3 \mu_3}{2} - 1} \exp\left( -\lambda_3 r^{\frac{\alpha_3}{2}} \right) \nonumber\\
	&~~~\times \left( \frac{\nu d_1^{m_1} d_2^{m_2} \sigma_R^2 \left( r + \frac{1}{\kappa}\right) }{P_S \left( 1 - \kappa \nu r \right)} \right)^{\frac{\alpha_2(\mu_1 + \mu_2)}{4}}  G_{1,3}^{2,1}\left( \left. \lambda_1 \lambda_2 \left( \frac{\nu d_1^{m_1} d_2^{m_2} \sigma_R^2 \left( r + \frac{1}{\kappa}\right)}{P_S \left( 1 - \kappa \nu r \right) } \right)^{\frac{\alpha_2}{2}} \right\vert \begin{array}{c}
	 1-\frac{\mu_1 + \mu_2}{2}\\
	 \frac{\mu_1 - \mu_2}{2}, -\frac{\mu_1 + \mu_2}{2}, \frac{\mu_2 - \mu_1}{2}
	 \end{array}\right) \textrm{d}r 
	\end{align}
	\hrulefill
\end{figure*}

\subsection{High SNR regime for the DF/AF-FD systems}
To the best of our knowledge, a closed-form solution does not exist for the integral given in \eqref{Pout_AF}. Therefore, we investigate the high SNR regime for the considered DF/AF-FD relaying systems.

\subsubsection{FD-DF} Using \eqref{gammaR_df} and \eqref{gammaD_df}, the effective SNR can be re-expressed as
\begin{align}
\lim\limits_{P_S \rightarrow \infty} \gamma_{\textrm{eff}}^{\textrm{DF}} = \min\left(\gamma_R,\gamma_D\right) = \gamma_R = \frac{1}{\kappa V},
\end{align} 
since $\gamma_R$ is independent of the transmit power $P_S$. 

\subsubsection{FD-AF} Considering \eqref{dest_snr_af}, the effective end-to-end SNR can be rewritten as
{\allowdisplaybreaks
\begin{align}
\lim\limits_{P_S \rightarrow \infty} \gamma_{\textrm{eff}}^{\textrm{AF}} = \frac{h_1^2 h_2^2}{\kappa h_1^2 h_2^2 h_3^2 + \cancelto{0}{\frac{d_1^{m_1} d_2^{m_2} \sigma_R^2}{P_S} \left(\frac{1}{\kappa} + h_3^2\right)}} = \frac{1}{\kappa V}.
\end{align}} 

With this in mind, at the high SNR regime, the OP metrics of both FD relaying systems coincide with each other and can be rewritten as follows
\begin{align}
\label{Pout_high}
\underset{P_S \rightarrow \infty}{P_{\textrm{out}}^{\textrm{DF/AF}}} = 1- \frac{\gamma_{inc}\left(\mu_3, \lambda_3 \frac{1}{(\kappa \nu)^{\alpha_3/2}} \right)}{\Gamma(\mu_3)}.
\end{align}
The corresponding results are presented in Fig. \ref{results3}.

\section{Numerical and Simulation Results}
In this section, we present analytical and numerical examples for the derived expressions. The adopted system parameters in our evaluations are as follows: $G=1$, $d_1 = d_2 = 5$ m, $m_1 = m_2 = 2$ W, $\sigma_R = \sigma_D = 0.01$ W and $\sigma_a = \sigma_c = \sigma_R/2$, $P_S = \{1; 10\}$ W, $\theta = 1$. By assigning different values for the $\alpha$ and $\mu$ parameters, we get the Rayleigh ($\alpha=2$ and $\mu=1$), Nakagami-$m$ ($\alpha=2$) and Weibull ($\mu=1$) fading coefficients.
 
Fig. \ref{results1} illustrates the analytical results for the OP as a function of $\mathcal{R}$ with various values of $P_S$ for both FD systems obtained using \eqref{Pout_DF1} and \eqref{Pout_AF}, respectively. For the sake of comparison and completeness, results for the HD systems are also appended to the figure, details can be found in \cite{tgcn}. It can be seen that the performance of FD schemes is vastly superior to that of the HD systems for a certain $\mathcal{R}$ irrespective of the value of $P_S$. It is also apparent that the DF systems always outperform those based on AF relaying due to the noise amplification which is specific to the AF relaying system. As expected, in all scenario, increasing the source power enhances the OP. 

\begin{figure}[!t]
	\centering
	\includegraphics[width=1\columnwidth]{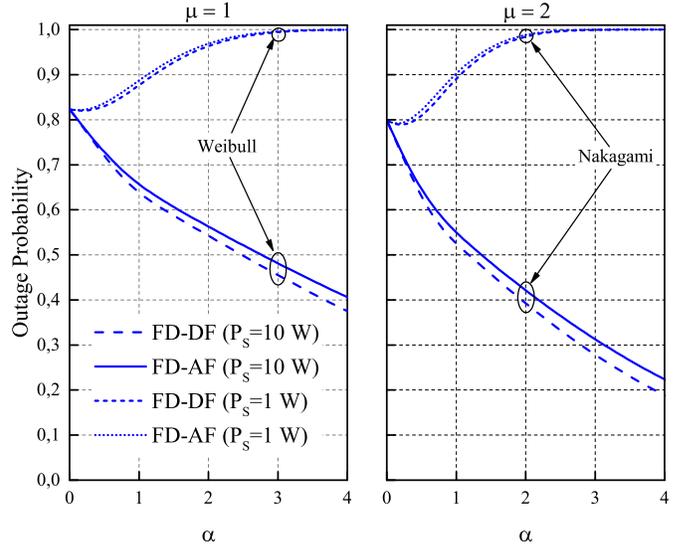}
	\caption{OP as a function of $\alpha$ for $\mu = \{1, 2\}$.}
	\label{results2}
\end{figure}

Now, to illustrate the impact of the fading parameter $\alpha$ on the system performance, we plot in Fig. \ref{results2} the OP for FD-DF and FD-AF systems versus $\alpha$. The increase of $\alpha$ improves the OP for $P_S = 10$ W and deteriorates performance when $P_S = 1$ W, respectively. This can be explained by the effective signal-to-LBI ratio obtained for different values of $P_S$. Similar observations can be seen when $\mu$ is varied. Note that the parameters $\alpha$ and $\mu$ are directly related to the power exponent and the number of multi-path components of the channel, respectively \cite{yacoub}. In addition, the corresponding values of the OP for various fading channels perfectly match the OP values shown in Fig. \ref{results1}. Therefore, the derived analytical expressions can be further applied for any other modifications of the fading channels derived from the i.n.i.d. $\alpha$-$\mu$ statistical model.

\begin{figure}[!t]
	\centering
	\includegraphics[width=1\columnwidth]{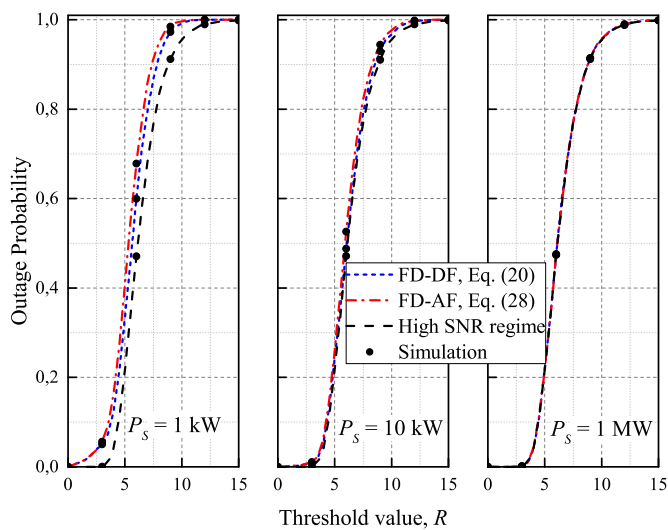}
	\caption{OP versus $\mathcal{R}$ for the FD systems with $P_S=\{1~\textrm{kW}, 10~\textrm{kW}, 1~\textrm{MW}\}$ over Rayleigh fading when $P_S \rightarrow \infty$.}
	\label{results3}
\end{figure}

Fig. \ref{results3} demonstrate some numerical results for the OP over Rayleigh fading as a function of $\mathcal{R}$ with various values of $P_S$ for the FD systems obtained from \eqref{Pout_DF1}, \eqref{Pout_AF} and \eqref{Pout_high}. It can be seen that the FD-DF system always outperforms the FD-AF even for high values of $P_S$, i.e., $P_S=\{1~\textrm{kW}, 10~\textrm{kW}, 1~\textrm{MW}\}$. As expected, in the high SNR regime, the OP metrics of the considered systems match each other.  

\section{Conclusion}
In this paper, we analyzed the OP of a two-hop TSR-based relaying system over i.n.i.d. $\alpha$-$\mu$ fading. Three special cases of the $\alpha$-$\mu$ fading channel, i.e., Rayleigh, Nakagami-$m$ and Weibull fading, were examined. FD was studied with both DF and AF relaying. We derived unified accurate analytical expressions for the OP which were validated with Monte Carlo simulations. The obtained results were compared with those for the HD systems. It was demonstrated that increasing the parameter $\alpha$ and/or $\mu$ improves the OP when the source transmit power is sufficient compared to the LBI level while deteriorating the performance when these power levels are found to be relatively equal. Finally, we analyzed the high SNR regime for the FD systems when $P_S \rightarrow \infty$. It was demonstrated that both AF and DF relaying systems achieve the same performance.

\section{Acknowledgment}
This publication was made possible by NPRP grant number 9-077-2-036 from the Qatar National Research Fund (a member of Qatar Foundation). The statements made herein are solely the responsibility of the authors.

\ifCLASSOPTIONcaptionsoff
\newpage
\fi

\ifCLASSOPTIONcaptionsoff
  \newpage
\fi

\end{document}